# Accelerated atomistic simulation study on the stability and mobility of carbon tri-interstitial cluster in cubic SiC

H. Jiang[1], C. Jiang[2], D. Morgan[1,2], I. Szlufarska[1,2]

[1]Materials Science Program, [2]Department of Materials Science & Engineering, University of Wisconsin, Madison, WI

**ABSTRACT**

Using a combination of kinetic Activation Relaxation Technique with empirical potential and *ab initio* based climbing image nudged elastic band method, we perform an extensive search of the migration and rotation paths of the most stable carbon tri-interstitial cluster in cubic SiC. Our research reveals paths with the lowest energy barriers to migration, rotation, and dissociation of the most stable cluster. The kinetic properties of the most stable cluster, including its mobility, rotation behavior at different temperatures and stability against high temperature annealing, are discussed based on the calculated transition barriers. In addition to fundamental insights, our study provides a methodology for investigation of other extended defects in a technologically important material.

Corresponding author: Izabela Szlufarska

Ph: (608) 265-5878

Fax: (608) 262-8353

Email: szlufarska@wisc.edu

1. **INTRODUCTION**

Silicon carbide is a wide-band-gap semiconductor material considered for high-power, high-frequency and high-temperature applications [1,2]. In its cubic form, SiC has been also proposed as a candidate structural material for next generation nuclear reactors due to the outstanding mechanical properties, good thermal stability and low neutron capture cross section of this material [3-5]. Stable defect clusters are known to form in SiC during ion implantation both in semiconductor applications [6] and in nuclear radiation environments [7,8]. These clusters have important consequences both for electronic properties and for radiation resistance of SiC [6-8].

Many efforts have been reported on the structure of stable defect clusters. For instance, Mattausch *et al.* [9-11] studied the Local Vibrational Modes (LVMs) of a series of carbon clusters consisting of a few interstitials and antisites, and proposed new candidates for the so-called P-U, P-T and $D_{II}$ photoluminescence centers; Jiang *et al.* [12,13] proposed a new ground state (GS) as well as a few low energy states of small carbon interstitial clusters with sizes up to 6 interstitials in cubic SiC. Bockstede *et al.* [10] investigated the annealing hierarchy of defects based on formation and binding energies of small defect clusters that had been reported in the literature. Defects clusters were assumed to act as aggregation centers of point defects in a low temperature regime (T < 1000 °C) and act as re-emitting sources of point defects by dissociation in a high temperature regime (T > 1000 °C). Up to this point, the mobility and dynamics of the defect clusters in SiC has not been investigated. Quantification of cluster dynamics is, however, important for understanding of processes that control high-temperature annealing of defects introduced during ion implantation and for predicting radiation resistance of SiC under given temperature and irradiation conditions. In addition to migration and dissociation of the clusters, cluster rotation is also of potential interest. For instance, recent studies of irradiation creep of SiC

[14,15] suggested that anisotropic distribution of small dislocation loops on {111} planes under the applied stress is responsible for the experimentally observed irradiation creep and swelling. These loops were hypothesized to be formed by self-interstitial clusters, and their formation and rotation behavior under stress is responsible for the anisotropic distribution. The authors characterized distribution of interstitial loops with sizes on the order of nanometers. It was found that while distribution of these loops was isotropic in the absence of stress, under tensile stress that distribution changed to anisotropic with about 20% more loops lying in planes perpendicular to the stress axis. The authors demonstrated that large (tens of nanometers sized) loops are not sufficient to explain the observed swelling and creep, and therefore the small (few nanometer and smaller sized) loops need to be accounted when trying to explain the aforementioned irradiation phenomena.

A major challenge in predicting dynamics of defect clusters in SiC lies in the high defect migration barriers in this material and in short simulation time scales of standard molecular dynamics (MD) simulations. Multiple accelerated techniques have been developed in the community to extend MD time scale limitations, such as hyperdynamics, parallel replica dynamics and temperature-accelerated dynamics developed by Voter *et al.* [16-18]. Another widely used technique for long-time scale simulation is kinetic Monte Carlo (kMC). In this technique, the system hops from one energy minimum to another based on the known transition probabilities. The time is advanced in each hop based on the transition state theory. However, kMC requires a predefined event lists for transitions out of each minimum, which is very challenging to achieve for cluster migration due to the unknown transitions and numerous intermediate states involved in cluster dynamics. To address this problem, several open-ended saddle point search algorithms have been proposed, including the activation-relaxation technique

(ART) [19-23], the dimer method [24], and the autonomous basin climbing method [25]. When the kMC is combined with one of these algorithms, an on-the-fly kMC scheme can be developed and it can work on systems with complicated energy surface for a long-time scale simulation.

In this paper, we employ the kinetic activation-relaxation (k-ART) technique to investigate dynamics of C interstitial clusters in SiC. Specifically, we focus on migration and rotation of the $(C_{BC})_3$ cluster, which has been proposed to be one of the most stable small interstitial clusters in irradiated SiC [13]. This cluster is composed of three carbon interstitials occupying 3 neighboring C-Si bond center sites in the {111} plane and it is a common building block of other small carbon interstitial clusters [13]. For example, the GS of carbon penta-interstitial cluster is composed of a $(C_{BC})_3$ cluster with a neighboring $(C_{BC})_2$ complex, and the GS of carbon hexa-interstitial cluster is composed of two neighboring $(C_{BC})_3$ clusters. The kinetic activation-relaxation technique (k-ART) based simulation protocol developed here for studies of the $(C_{BC})_3$ cluster can be later employed to study other defects in this material.

## 2. METHODS

### 2.1. k-ART sampling

The rotation and migration paths of the $(C_{BC})_3$ cluster are investigated using ART, a single-ended eigenvector-following method developed by Mousseau *et al.* [19] and modified later by several groups [26-28]. In the activation phase, the system is pushed out of equilibrium by displacing selected atoms in steps of length 0.1 Å in random directions of the configuration space and a limited energy minimization is applied in the orthogonal hyperplane after each step. Selected atoms can be picked up randomly in the system or limited to, for example, the surface of a slab or a region containing defects. In our case, in order to focus on the movement of the interstitial cluster, selected atoms are chosen to be the interstitials in the cluster and their nearest neighbor lattice atoms. The system is moved step by step until a configuration is found where the Hessian matrix has a negative eigenvalue. In the convergence phase, the system is moved along the eigenvector of the negative eigenvalue of the Hessian matrix with an adaptive step length until either a saddle point is found or the system falls back into the original local minimum. In the former case, the system is subsequently pushed over the saddle point and relaxed into a local minimum. In the latter case, a new random search is started from the initial configuration. Although ART can work with *ab initio* software, e.g., with SIESTA and BigDFT [22], the large number of force evaluations required to converge to a saddle point makes such calculations very time consuming. In our simulations, we combine ART with the environment-dependent interatomic potential (EDIP) [29]. The EDIP potential is chosen since it provides a good description of energetics of point and extended defects [12,13], which is particularly important to our study. The key EDIP predictions are then further assessed with first-principles methods. The ART algorithm is combined with the kMC scheme in such a way that ART samples nearby saddle points of a local minimum to generate event lists with corresponding rates and the kMC algorithm executes an event according to these rates. This approach is referred as the k-ART and has been described in detail in Ref. [23]. To boost the efficiency of k-ART in sampling nearby saddle points, we parallelize the code with the message passing interface (MPI) so that several processors work on each selected atom for a certain time and then report all generated events to a master processor to obtain the complete event list.

We performed a few k-ART simulations on the $(C_{BC})_3$ cluster at 1500 K and no migration is observed even after hundreds of thousand kMC steps (corresponding to 0.1 µs). This result implies that the migration barrier of the cluster is high. To activate the migration of the cluster on the timescale of the kMC simulations, we raise the simulation temperature to 5000 K. It is important to mention that this high temperature only affects the rate of events associated with cluster dynamics and, for example, it does not cause melting of SiC. The limited impact of the high temperature is because the crystal is relaxed to a local minimum in each kMC step, which effectively corresponds to modeling the crystal at 0 K. One problem with the high simulation temperature is that the cluster can dissociate as well as migrate in the kMC simulations. Since we want to isolate the migration from the dissociation of the cluster, in the kMC algorithm we apply a constraint so that dissociation effects are excluded. Dissociation events are identified as those where the largest distance between selected carbon atoms in the cluster and its nearest neighbor atoms is larger than one and a half lattice constants. We also apply a stopping criterion to stop the simulation when the system completes a migration or a rotation, which are identified as the system is in a state with energy difference from GS smaller than 0.25eV and with a displacement from the initial configuration larger than a specified threshold value. The displacement threshold can be set to be small (~4 Å) to find rotation pathways or to be large (~6.5 Å) to force the cluster to migrate. With these modifications, k-ART can identify migration or rotation paths within few hundreds of kMC steps. Dissociation is considered separately later in this paper.

### 2.2. Ab initio calculations

The energy surface of paths predicted by k-ART is refined by density functional theory (DFT) calculations with the climbing image nudged elastic band (CI-NEB) method [30], as implemented in the Vienna Ab initio Simulation Package (VASP) [31]. The electron-ion interactions are described by the projector-augmented wave (PAW) method. We use PAW pseudopotentials with the valence electron configurations of $2s^2$, $2p^2$ for C and $3s^2$, $3p^2$ for Si. The cutoff energy for plane-wave basis sets is set at 500 eV. For Brillouin zone sampling, a 2×2×2 Monkhorst-Pack k-point mesh is used. We have tested the supercell size convergence on a cluster rotation barrier. We find that the error of the transition barrier within a 4×4×4 supercell

(512 atoms) is converged to approximately 50 meV when compared with the barrier extrapolated at infinite size, while the error of the same transition barrier within a 3×3×3 supercell (216 atoms) is only converged to 0.12 eV. Therefore, a 4×4×4 supercell should be used when refining barriers predicted from EDIP. Considering the efficiency of DFT calculations, we first calculate the energy surface of each path with a 3×3×3 supercell, and then refine the transition step with the highest barrier in a 4×4×4 supercell. Jiang *et al.*[12] have shown that the $(C_{BC})_3$ structure is neutral for most values of the Fermi level, so we consider only neutral interstitials in all of our DFT calculations. By computing the quantum mechanical Hellmann-Feynman forces, all internal atomic positions are fully optimized using quasi-Newton method for CI-NEB images and conjugate gradient method for end points until forces are less than 0.02 eV/Å.

3. **RESULTS**

The two most stable configurations of the $(C_{BC})_3$ cluster in the cubic SiC are shown in Figure 1 and one can see that these clusters lie within the {111} planes. A C-centered (or Si-centered) cell refers to the cell where the lattice atom right above the center of the $(C_{BC})_3$ cluster is C (or Si). The C- and Si- centered cells are shown in Figure 1(a) and Figure 1(b), respectively. Jiang *et al.* [12] showed that the $(C_{BC})_3$ cluster within a C-centered cell is the GS of carbon tri-interstitial clusters in cubic SiC, and the energy of that within a Si-centered cell is 0.47 eV higher than GS in DFT, and 0.25 eV higher in EDIP. Thus, a stopping criterion of 0.25 eV within EDIP is sufficient to include both the C-centered and Si-centered structure. All of our k-ART simulations start from the C-centered structure. We define the rotation of the cluster as the movement from one {111} plane to another {111} plane but within the same C-centered cell, and migration as the movement from one C-centered cell to another C-centered cell either directly or through a Si-centered cell.

An example of energies calculated in a successful k-ART simulation as a function of the kMC step is shown in Figure 2(a). The displacement from the initial configuration is calculated as

$$r = (\sum_{i=1}^{N} \Delta r_i^2)^{1/2} \qquad (1)$$

where $\Delta r_i$ is the displacement of atom $i$ and $N$ is the total number of atoms in the entire simulation box. The high simulation temperature drives the system to explore the potential energy surface aggressively by overcoming saddle points with energies up to 3 eV, which are unlikely to occur at low temperature simulations. We introduce a displacement threshold of 4.0 Å, which means that when the system is in a state with displacement from the initial C-centered structure larger than this threshold and with energy difference from GS less than 0.25eV, a migration of rotation path is completed and the simulations is stopped. The value of the displacement threshold has been varied during the sampling between 4.0 Å and 6.5 Å in order to force the cluster to move to different sites. For each successful simulation, we visualize the configurations along the identified path. In some cases, the system does not find the GS after thousands of kMC steps, and such simulations are labeled as failed since it yields neither a migration nor a rotation path. These failed simulations usually run into states with the relative energy higher than 8 eV because of the high simulation temperature. As shown in Figure. 2(b), we obtain 68 successful simulations out of the total of 131 simulations, and the 68 successful simulations yield 12 distinct migration paths and 4 distinct rotation paths. The number of distinct path starts at 7 because we have identified 7 paths when running tests on the codes. After finding no new paths for the last 45 successful simulations, the search is considered exhaustive. Subsequently all of the identified migration and rotation paths are refined by CI-NEB method based on DFT calculations.

## 3.1. Migration and dissociation

In Table I, we show the energy barriers of the 12 distinct migration paths calculated within EDIP and DFT. A detailed description of the energy surface and intermediate configurations along each path shown in Table I can be found in the supplemental materials. The ID for each path is assigned according to their similarity in migration mechanism. For example, all paths with ID starting as M2 represent a process of rotation of the cluster in the original (111) plane. Path M7 turned out to be the same as path M4C after refining it with DFT calculations, so its barrier within DFT is labeled as M4C. Most of the barriers predicted by EDIP lie between 3.0 eV and 4.5 eV, while the DFT refined barriers lie between 4.0 eV and 5.5 eV. Here we focus on the lowest barrier paths predicted by EDIP and refined by DFT.

As can be seen in Table I, the path with the lowest migration barrier as determined from DFT is path M1 with the barrier of 4.29 eV in DFT and the barrier of 4.12 eV in EDIP. The energy surface along this path and intermediate configurations are shown in Figure 3. In this path, one interstitial first rotates around its nearest neighbor Si atom to form a dumbbell with the C lattice atom that used to occupy the position immediately above the $(C_{BC})_3$ cluster in its GS configuration (configuration α in Fig. 3). In the next step, the same C atom moves between the other two interstitials (configuration β) and enters the neighboring Si-centered cell (configuration γ). Finally, that C atom rotates around its nearest neighbor C lattice atom to recover the $(C_{BC})_3$ structure within a neighboring Si-centered cell in the original (111) plane. Along this path, the cluster moves from one C-centered cell to its neighboring Si-centered cell, and it can move to any neighboring C-centered cell from this Si-centered cell by inversing this path.

It is instructive to discuss the M6 migration path, which was found to have the lowest migration barrier (2.84 eV) in EDIP. The corresponding barrier in DFT is 4.37 eV. The energy surface and intermediate configurations for this path are shown in Figure 4. In this path, one of the interstitial first rotates around its nearest neighbor Si lattice atom to form a C-C dumbbell with the C atom that used to occupy the center position above the cluster in its GS configuration (configuration α in Fig. 4). In the next step (β), another interstitial rotates around its nearest neighbor C lattice atom, leaving the initially occupied C-Si bond center site vacant. This vacant C-Si bond center site is being simultaneously filled by the first C interstitial that moves towards it (γ). As a result, a tri-angular bonded structure composed of 3 C atoms at a C lattice site formed. In the next step (δ), one of the interstitials moves into a neighboring C-centered cell and thereby a chain of six C atoms is formed. The chain has C-C dumbbells at both ends and these dumbbells can rotate leading to a symmetry equivalent configuration (δ'). After that, the tri-angular bonded structure (γ') is formed again, this time in a new C-centered cell as the interstitial near the end of the chain moves from the original C-centered cell to the new cell. Since γ' and C-centered' are symmetry equivalent to γ and C-centered GS, respectively, the cluster can transform from γ' to C-centered' by reversing the pathway from C-centered to γ, as described above. Following this path, a C lattice atom is pushed to the bond center site while the vacant lattice site is now filled by an initial C interstitial.

We define dissociation as the process during which the $(C_{BC})_3$ cluster breaks into two non-interacting defects: a single C interstitial and a C di-interstitial defect. The final configuration of the single C interstitial is chosen to be the C split interstitial in the tilted <100> configuration ($C_{sp<100>}$). The final state of the C di-interstitial defect is $(C_{BC})_2$, where two C interstitials occupy neighboring C-Si bond center sites in (111) plane. These two final structures are chosen because they are the most stable configurations for the respective defects in cubic SiC [13,32]. In fact, $(C_{BC})_3$ and $(C_{BC})_2$ share the same bond center bonding structure with the only difference that $(C_{BC})_3$ is composed of 3 bond centers bonding to each other while $(C_{BC})_2$ composed of only 2 bond centers. This implies that the dissociation is a process that one C interstitial in $(C_{BC})_3$ breaks bonds with the other two and migrates away until there is no interaction between it and the $(C_{BC})_2$ left behind. The energy of the final dissociate state can be calculated by summing up formation energies of the non-interacting constituents and it is equal to 4.36 eV. The high symmetry of $(C_{BC})_3$ limits the number of possible dissociation pathways to a few, and after calculating them with DFT, we determined the lowest dissociation barrier as 4.83 eV, as shown in Figure 5. Along this pathway, one of the interstitial in $(C_{BC})_3$ first breaks bonds with the other two and then migrates away from the $(C_{BC})_2$ by C-C/C-Si dumbbell transition mechanism, as suggested in Ref. [32]. Note that along this pathway, the relative energy visits the 4.36 eV state twice, however, we take the higher barrier (4.83 eV) which leads to the second 4.36 eV state as dissociation barrier. This barrier relative to the dissociation state is 0.47 eV, which agrees with the 0.5 eV migration barrier of a single $C_{sp<100>}$ [32,33], and this implies there is no interaction between $C_{sp<100>}$ and $(C_{BC})_2$. This is not true for the first 4.36 eV state, where the migration barrier of the $C_{sp<100>}$ is lowered due to the interaction between the two defects.

Given the above migration and dissociation barriers, the $(C_{BC})_3$ cluster can be viewed as immobile and stable at temperatures below 1100 K, where neither migration nor dissociation can take place on long-time scale (hop rate < 1 month$^{-1}$). The hop rate is calculated by using the following equation,

$$\Gamma = v e^{(-E_a/kT)} \quad (2)$$

where $\Gamma$ is the hop rate, $v$ is the attempt frequency approximated as $5 \times 10^{12}$ s$^{-1}$, $E_a$ is either the migration or dissociation barrier, $k$ is the Boltzmann constant, and $T$ is the absolute temperature. Therefore, in the low temperature regime (T < 1100 K) this defect is likely to serve as aggregation center for interstitials that then grow into larger clusters or planar defects as proposed by Jiang *et al.*[13]. It is also interesting to consider the long term evolution of the $(C_{BC})_3$ defect at higher temperatures in typical annealing experiments (over 1300 K). We define a critical temperature of dissociation as the temperature at which dissociation can take place once within the time of typical annealing experiment (~1 hr [34]). To estimate this temperature, we use equation (2) and set $\Gamma$, the dissociation rate, to be once per the experimental time of 1hr. The critical temperature of dissociation is then calculated as 1500 K based on the barrier of $E_a = 4.83$ eV. This analysis implies that, at temperatures below 1500 K, annealing of $(C_{BC})_3$ defects will be dominated by their migration whereas dissociation begins to be dominate at temperatures higher than 1500 K. It is useful to quantify the mobility of the $(C_{BC})_3$ cluster in the regime where mobility dominates the annealing. To do that, we calculate the mean diffusion distance $x$ for the experimental time scale $\tau = 1$ hr at a given temperature $T = 1500$ K using the following equation

$$D = \frac{x^2}{2d\tau} = a^2 v e^{(-E_m/kT)} \qquad (3)$$

where $D$ is the diffusion coefficient, $d$ is the dimensionality of the system, $a$ is the hop distance, $E_m$ is the migration barrier. The dimensionality $d$ is taken as 3 because the rotation of the cluster among different (111) planes (discussed later in Section 3.2) can be activated at 1500 K. The hop distance $a$ is taken as the displacement of the center of mass of the $(C_{BC})_3$ defect in a migration between two neighboring C-centered cells following path M1, and is calculated as 3.1 Å. Migration barrier $E_m$ is taken as 4.29 eV, the lowest barrier of migration path (M1). This analysis predicts the mean diffusion distance $x \approx 6.3$ nm, which is much smaller than the grain size of typical experimental SiC samples. Therefore, the mobility of the $(C_{BC})_3$ cluster is very low even at 1500 K, and cannot be easily annealed out by diffusion to sinks alone.

Because the $(C_{BC})_3$ structure is a common build block of small carbon interstitial clusters in cubic SiC [13], based on the knowledge of its mobility and stability, a reasonable guess on the kinetics of small carbon interstitial clusters can be made. First, we assume the migration barrier of small carbon interstitial clusters is comparable to the migration energy barrier of 4.29 eV found for the $(C_{BC})_3$ cluster. Secondly, we approximate the dissociation barrier with the following equation

$$E_{dissociation}(n) = E_b(n) - E_b(n-1) + E_{M,Ci} \qquad (4)$$

where $E_b(n)$ is the binding energy of carbon $n$-interstitial cluster, defined as the energy difference between a carbon $n$-interstitial cluster and $n$ non-interacting $C_{sp<100>}$, and $E_{M,Ci}$ is the migration barrier of a single C interstitial taken as 0.5 eV, based on Refs. [32,33]. In Table II, we show the binding energies of small carbon interstitial clusters with a size of up to 6 and their approximated dissociation barriers. For clusters with size over 3, the dissociation barrier is around 4 eV, which is comparable to or even lower than the cluster migration barriers. Therefore, in high temperature annealing, small carbon interstitial clusters will likely dissociate into smaller clusters by emitting interstitials one by one with limited or no diffusion of the clusters themselves. It would be safe to assume that small carbon interstitial clusters with sizes equal to or larger than 3 in cubic SiC are immobile in long-time scale modeling of defect evolution, such as cluster dynamics model. However, one exception is the $(C_{BC})_2$ defect, the dissociation barrier of which is approximated as 5.66 eV. Since the migration barrier of this defect is not higher than 4.29 eV, it is diffusive at temperatures below the critical dissociation temperature and therefore the dynamics of $(C_{BC})_2$ defect should be carefully treated in long-time scale simulations.

Based on the good agreement between calculated and experimental LVMs, Jiang et al. [12] proposed that the $(C_{BC})_3$ cluster as a candidate for the $D_{II}$ center in SiC, which has been found in photoluminescence experiments [35-39]. Up to this point, the $(C_{BC})_3$ cluster was found to be the only candidate that produces 5 LVMs in the cubic SiC and 10 LVMs in the hexagonal SiC, in agreement with experiments, and one that provides the best (although not excellent) quantitative agreement with the LVMs. These findings support the idea that the $(C_{BC})_3$ cluster is responsible for the experimentally observed $D_{II}$ center. We are now in the position to ask if the thermodynamic stability of $(C_{BC})_3$ is consistent with some experimental studies that found that the defect responsible for the $D_{II}$ center can persist even after annealing at 1973 K [35-37]. Our calculated migration and dissociation energies of $(C_{BC})_3$ are inconsistent with this proposal. This inconsistence may arise from multiple factors. First of all, there are limited details on experiments reported in literature, which makes the comparison somewhat qualitative. So far, only one paper from year 1973 [35] reported the persistence of $D_{II}$ center up to 1973 K in cubic SiC and this paper gives very little details on the doping conditions, temperature fluctuation and annealing time. Other papers [36,37] that reported the persistence of $D_{II}$ center up to 1973 K were based on experiments conducted with hexagonal (4H) SiC samples, which may possibly change the dissociation barrier. To make comparison to experiments conducted with 4H-SiC samples, we calculate the dissociation energy of the $(C_{BC})_3$ cluster in 4H-SiC and approximate the dissociation barrier by adding the migration barrier (0.5 eV) to the dissociation energy. The approximated dissociation barrier is 5.0 eV, which is still not high enough for the defect to survive during 1973 K annealing (hop rate is ~1 s$^{-1}$). Another possibility to explain the lack of stability of $(C_{BC})_3$ at 1973 K is that this cluster is stabilized against annealing by binding to another defect, such as an antisite or Si interstitial. It is also possible that the defects responsible for the $D_{II}$ center form other configurations that are derived from the three-bond center structure as in the $(C_{BC})_3$ cluster, for example by forming larger clusters that grew by incorporating Si interstitials. Such structure might share similar LVMs and stability with the $(C_{BC})_3$ but have a high barrier to dissociation. Finally, it is possible that the $(C_{BC})_3$ cluster is reassembled in the experimental samples when they are quenched from high temperature to 0 K, at which the photoluminescence tests were conducted. The high mobility of C interstitial increases the probability that point defects find each other and are stabilized as $(C_{BC})_3$ during quenching below the dissociation temperature.

*3.2. Rotation*

As can be seen in Table III, the discrepancy in energy barriers for rotation paths between EDIP and DFT is also non-negligible. However, the actual rotation path corresponding to the lowest energy barrier predicted by EDIP (with the energy barrier of 1.84 eV) agrees with the lowest path predicted after DFT refinement (with the energy barrier of 4.14 eV). The energy surface and intermediate states of the lowest barrier path are shown in Figure 6. In this pathway, one C interstitial first rotates around its nearest neighbor Si lattice atom to form a C-C dumbbell with the C lattice atom (configuration α), which used to occupy the center position right above the $(C_{BC})_3$ cluster in its GS. Next, another C interstitial rotates around its nearest neighbor C lattice atom and moves to the (111) plane that is already occupied by the first interstitial (configuration β). Finally, the last C interstitial moves to the same (111) plane by rotating around its nearest neighbor C lattice atom and the GS $(C_{BC})_3$ structure is recovered.

With a transition barrier of 4.14 eV, the rotation of the $(C_{BC})_3$ cluster between (111) planes can barely happen at temperatures below 1100 K (less than 1 hop rate per month). This implies that once the $(C_{BC})_3$ cluster is formed on a certain (111) plane, it can grow into larger clusters on the same (111) plane, consistently with the idea that this cluster is an aggregation center for interstitials. Rotation of the cluster can be activated at elevated temperatures with a rate comparable to that of migration (the lowest barrier to migration is 4.29 eV).

## CONCLUSIONS

Using a combination of k-ART sampling with EDIP and *ab initio* based CI-NEB calculations, we have determined the migration and dissociation energies of the $(C_{BC})_3$ cluster in cubic SiC, which is the most stable small C interstitial clusters in SiC among the known clusters in SiC. The fastest migration path has the energy barrier of 4.29 eV and the second fastest path has a comparable barrier of 4.37 eV. The dissociation barrier of the $(C_{BC})_3$ cluster into a non-interacting $(C_{BC})_2$ defect and $C_{sp<100>}$ is 4.83 eV and the rotation barrier of the cluster moving between different (111) planes is found to be 4.14 eV. The mobility of the $(C_{BC})_3$ cluster is limited up to temperatures of ~1500 K, and thus this cluster can be treated approximately as immobile. The $(C_{BC})_3$ cluster was found to dissociate as temperature above 1500 K and it is predicted that this is the pathway by which it can be annealed out from the crystal, rather than through diffusion to a sink. Since the $(C_{BC})_3$ structure is a common building block of small carbon interstitial clusters, it is reasonable to assume their lack of mobility in long-time scale modeling such as cluster dynamics. An interesting direction for the future study will be the effects of applied stress on the rotation of small clusters and large dislocation loops in SiC, since it had been proposed that rotation of such defects and the resulting anisotropic distribution of dislocation loops with respect to the stress axis are responsible for irradiation swelling and creep in SiC.[14,15]

## ACKNOWLEDGEMENT

This research is supported by the US Department of Energy, Office of Basic Energy Sciences Grant No. DE-FG02-08ER46493. This work used the Extreme Science and Engineering Discovery Environment (XSEDE), which is supported by National Science Foundation grant number OCI-1053575.

**Figures and Tables**

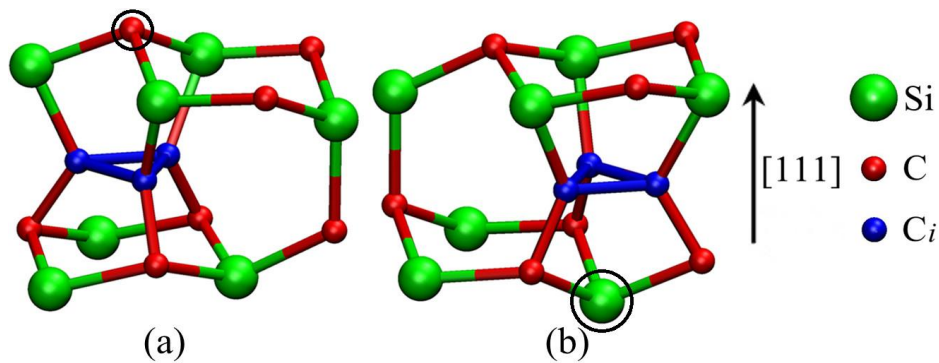

**FIG.** 1. (Color online) Two possible cells in which the $(C_{BC})_3$ cluster can reside in cubic SiC: (a) C-centered cell; (b) Si–centered cell. The black circles label the centered C and Si atoms, respectively.

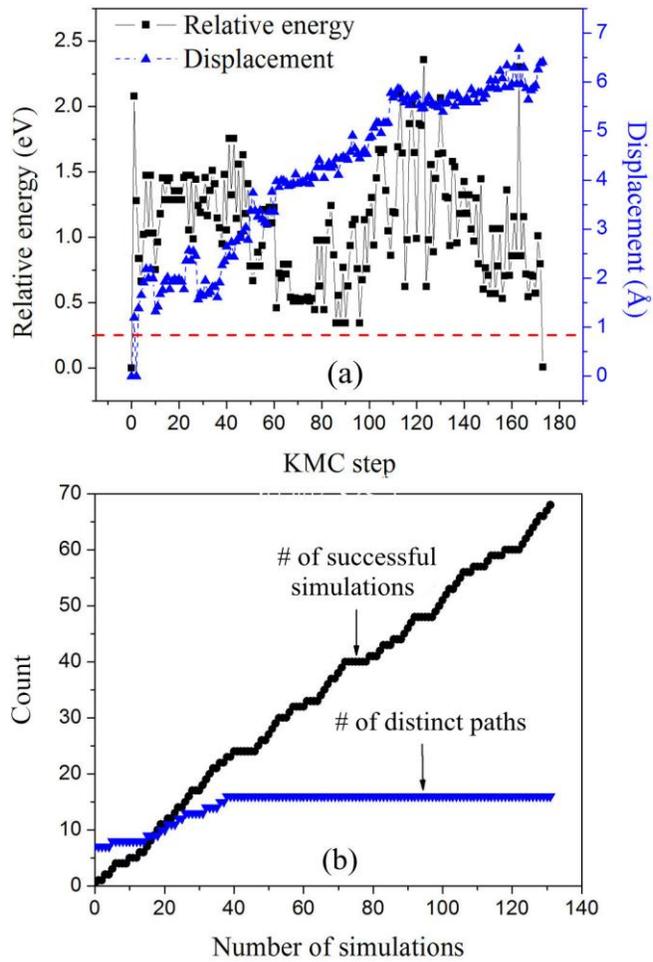

**FIG.** 2. (Color online) (a) Evolution of the energy and displacement from the initial C-centered configuration as a function of kMC steps in one successful k-ART simulation. The energy is calculated relatively to the initial configuration. The displacement is defined in Eq. (1) in text. The horizontal dashed line (red online) represents the stopping criterion of 0.25 eV. (b) The number of successful simulations and distinct paths identified from these successful simulations.

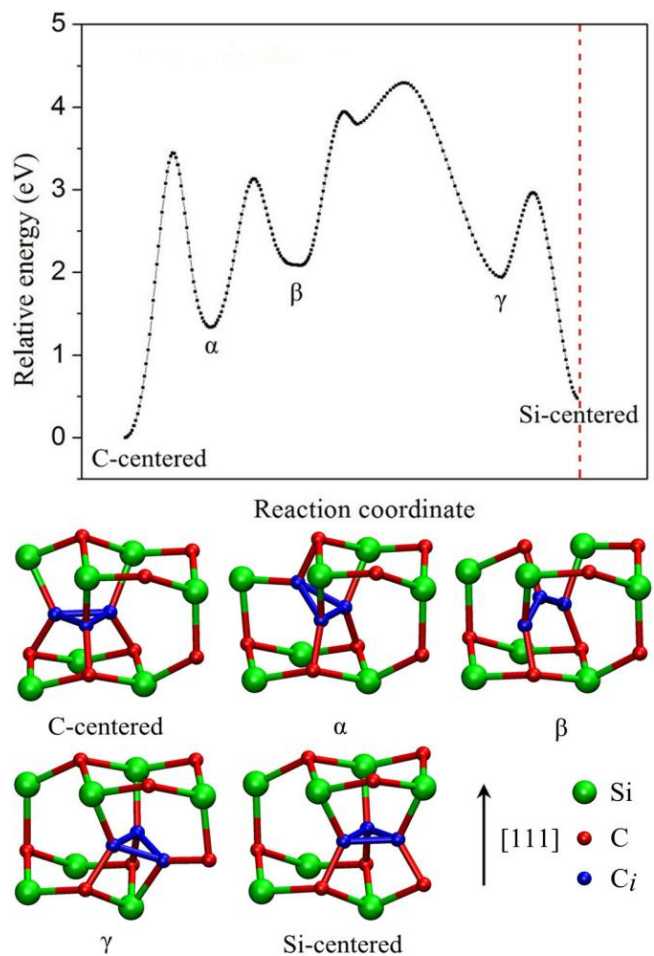

**FIG.** 3. (Color online) Schematic view of the energy surface and of the intermediate states along path M1. The energy surface is symmetric with respect to the vertical dashed line (red online). α, β and γ are local minima along the pathway.

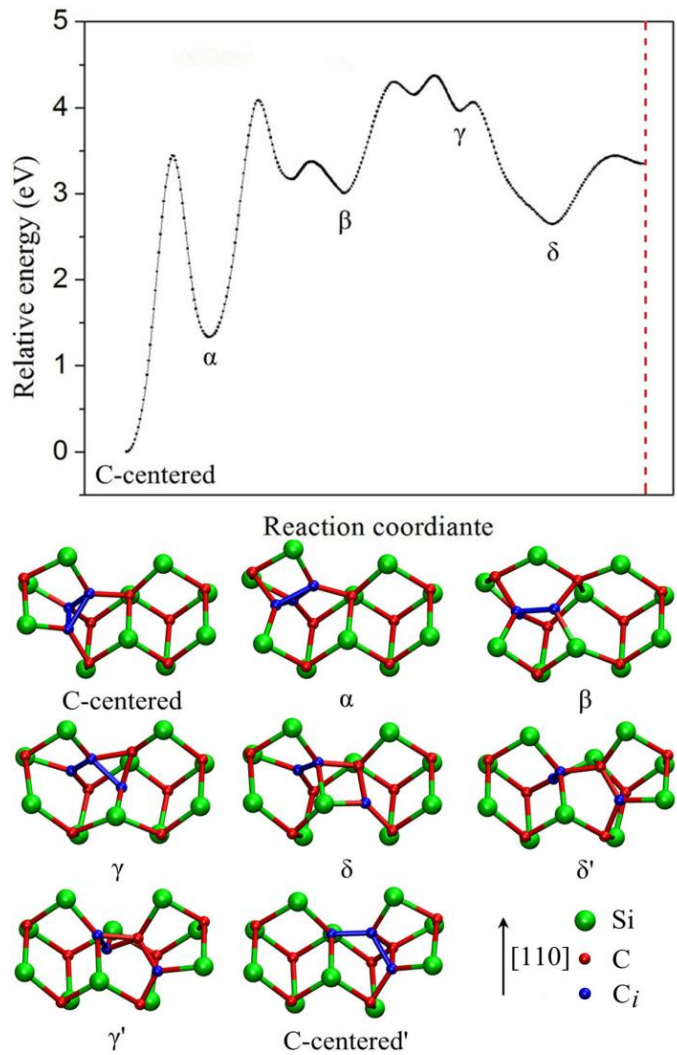

**FIG.** 4. (Color online) Schematic view of the energy surface and of the intermediate states along path M6. The energy surface is symmetric with respect to the vertical dashed line (red online). α, β, γ and δ are intermediate local minima along the pathway, γ', δ' and C-centered' are symmetry equivalent configurations of γ, δ and C-centered GS, respectively, in another C-centered cell. Only the three initial C interstitials are labeled as blue sphere.

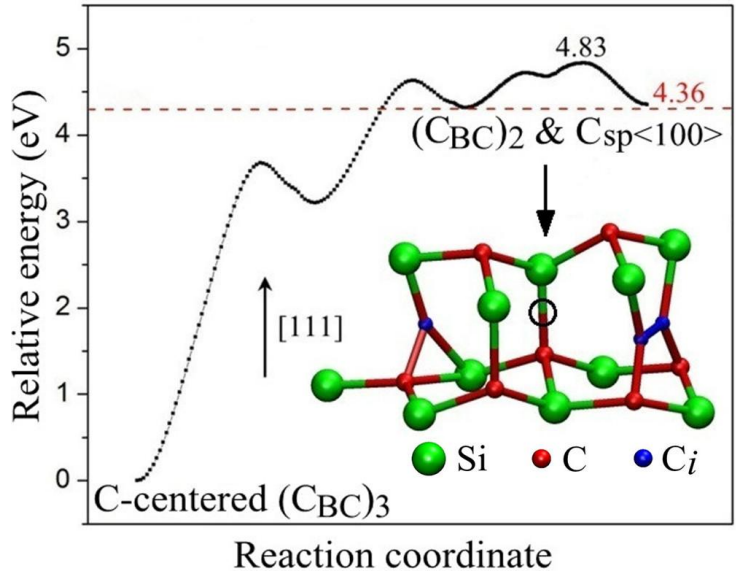

**FIG.** 5. (Color online) The energy surface for the dissociation of the $(C_{BC})_3$ cluster. The dashed horizontal line (red online) at 4.36 eV represents the relative energy of not-interacting $(C_{BC})_2$ defect and $C_{sp<100>}$. The embedded structure represents the final configuration, where the solid circle represents the initial bond center site occupied by the $C_{sp<100>}$.

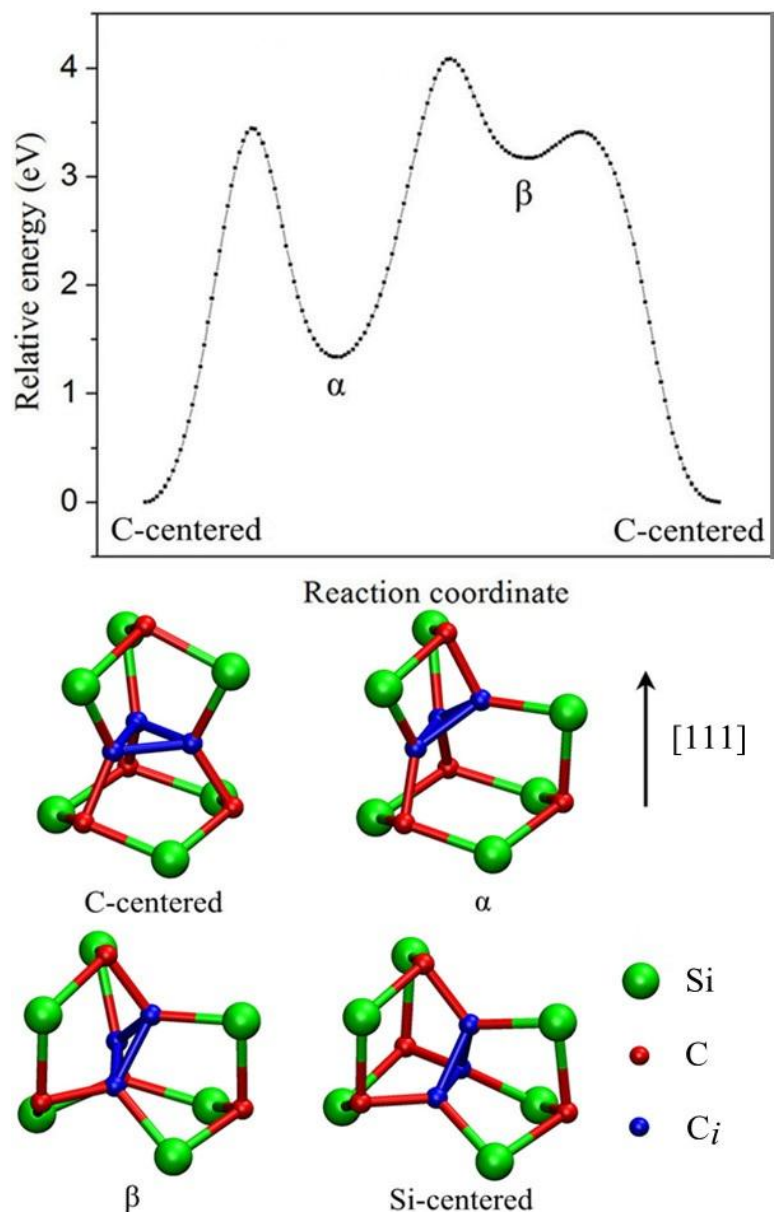

**FIG.** 6. (Color online) Schematic view of the energy surface and of the intermediate states of the rotation path R1. α and β are intermediate local minima along the pathway.

**TABLE** I. Energy barriers calculated with EDIP and DFT for the 12 migration paths identified with EDIP.

| Path ID | Energy barrier (eV) | |
|---|---|---|
| | EDIP | DFT |
| M1 | 4.12 | 4.29 |
| M2A | 3.09 | 5.26 |
| M2B | 4.53 | 5.26 |
| M2C | 5.42 | 5.24 |
| M3A | 3.03 | 5.40 |
| M3B | 3.47 | 5.40 |
| M4A | 3.22 | 4.77 |
| M4B | 4.13 | 7.51 |
| M4C | 3.04 | 5.31 |
| M5 | 4.32 | 4.99 |
| M6 | 2.84 | 4.37 |
| M7 | 3.21 | M4C |

**TABLE** II. Binding energies and approximated dissociation barriers of the GS of small carbon interstitial clusters with size up to 6. Binding energy is taken from Ref. [13]. $(C_{BC})_{3,C}$ represents a $(C_{BC})_3$ structure in the C-centered cell, and $(C_{BC})_{3,Si}$ represents a $(C_{BC})_3$ structure in the Si-centered cell.

| Size | Structure | $E_b$(eV) | $E_{dissociation}$(eV) |
|---|---|---|---|
| 2 | $(C_{BC})_2$ | 5.16 | 5.66 |
| 3 | $(C_{BC})_{3,C}$ | 9.52 | 4.86 |
| 4 | $(C_{sp})_4$ | 12.37 | 3.35 |
| 5 | $(C_{BC})_{3,C}+(C_{BC})_2$ | 16.12 | 4.25 |
| 6 | $(C_{BC})_{3,C}+(C_{BC})_{3,Si}$ | 19.46 | 3.84 |

TABLE III. Energy barriers of the 4 rotation paths found in EDIP and calculated within EDIP and DFT.

| Path ID | Energy barrier (eV) | |
|---|---|---|
| | EDIP | DFT |
| R1 | 1.84 | 4.14 |
| R2 | 4.13 | 6.82 |
| R3 | 3.52 | R1 |
| R4 | 3.72 | 5.70 |